\documentclass[aps,prl,amsmath,amssymb,twocolumn]{revtex4}
\usepackage{amssymb}
\usepackage{graphicx}
\begin{document}

\title{Landscapes and Fragilities}

\author{G.~Ruocco$^1$, F.~Sciortino$^{1,2}$,
F.~Zamponi$^1$, C.~De~Michele$^{1}$, T.~Scopigno$^1$}

\affiliation{$^1$ INFM and Dipartimento di Fisica, Universit\'a di Roma "La
Sapienza", 00185 Roma, Italy \\ $^2$ INFM Center for Statistical Mechanics and
Complexity}

\date{\today}
\begin{abstract}
The concept of fragility provides a possibility to rank different
supercooled liquids on the basis of the temperature dependence of
dynamic and/or thermodynamic quantities. We recall here the
definitions of kinetic and thermodynamic fragility proposed in the
last years and discuss their interrelations.  At the same time we
analyze some recently introduced models for the statistical
properties of the potential energy landscape. Building on the
Adam-Gibbs relation, which connects structural relaxation times to
configurational entropy, we analyze the relation between
statistical properties of the landscape and fragility. We call
attention to the fact that the knowledge of number, energy depth
and shape of the basins of the potential energy landscape may not
be sufficient for predicting fragility. Finally, we discuss two
different possibilities for generating strong behavior.
\end{abstract}
\maketitle

\section{Introduction}
Soon after the introduction of the concept of "topographic view of
the Potential Energy Landscape (PEL)"
\cite{Goldstein69,Stillinger84}, it became immediately clear that
a key role in controlling the kinetic arrest characterizing the
glass transition was played by the number of distinct \cite{nnn}
PEL local minima (inherent structures), $\Omega_N$, and by their
energy distribution, $\Omega_N(E)$. Indeed, it was suggested that
the qualitatively different behavior of different supercooled
liquids could be traced back to the difference in the $\Omega_N$
function, or, more specifically to the steepness of the $N$
dependence of this quantity. From general arguments, in a
mono-component collection of a large number, $N$, of units (atoms,
molecules, ...), it can be shown that $\Omega_N \sim 
\exp{(\alpha N)}$. Similarly, it holds $\Omega_N(E)\sim
\exp{(\Sigma(e)/k_B)}$. Here $\Sigma(e)$ assumes the meaning of
"configurational entropy" and it is an extensive function of the
energy per particle $e=E/N$. The quantity $\alpha$
($\alpha=max_e\{\Sigma(e)\}/Nk_B$) is a measure of the total
number of "inherent structure" (individual minima of the potential
energy hyper-surface). In comparing the behavior of different
glass forming systems, particular emphasis is placed in the
relation existing between $\alpha$ and the "fragility" of the
system under investigation.

The "fragility" concept, in its modern form, 
has been introduced, developed and widespread by Angell
\cite{Angell85}. It describes, in its kinetic version, how fast
the structural relaxation time ($\tau_\alpha$) increases with
decreasing temperature on approaching the glass transition
temperature, $T_g$, defined as the temperature where $\tau_\alpha$
becomes equal to 100 s. "Strong" systems (low values of fragility)
show a "weak" $T$-dependence of $\tau_\alpha(T)$, which can be
described by an Arrhenius law ($\tau_\alpha(T)=\tau_\infty
\exp{(\Delta / k_B T)}$) while "Fragile" systems show -close to
$T_g$- a much faster $T$ dependence of the relaxation time, which
is also markedly non-Arrhenius (this dependence could be, for
example, described by a $T$- dependence of the activation energy
$\Delta$). The relaxation time is a quantity which is rather
difficult to access, in particular when the value of $\tau_\alpha$
is large, and, moreover, it seems also to be technique-dependent.
For these reasons, in non-polymeric liquids, the fragility is
usually defined through the $T$ dependence of the shear viscosity,
$\eta$ \cite{Angell91}. This choice leads to a first ambiguity,
especially in comparing different systems, as the fragility
defined through $\tau_\alpha(T)$ and that defined through
$\eta(T)$ are not coincident. This can be rationalized by
recalling the Maxwell relation, $\eta$=$G_\infty \tau_\alpha$
(here $G_\infty$ is the infinite frequency shear modulus of the
liquid), and recalling that $G_\infty$ at $T_g$ spans over about
two decades among different systems. Another possible definition
of fragility comes from the temperature dependence of the mass
diffusion coefficient. In this case, according to the
Stokes-Einstein relation ($D=k_BT/(6\pi r \eta)$, being $r$ the
effective hydrodynamic radius), it is the {\it mobility} $\mu$
(=$D/T$) that is (inversely) proportional to the viscosity and,
therefore, must be analyzed. Once more, it should be expected that
the fragility defined via mobility and that defined via viscosity
are not coincident. Indeed, i) the effective hydrodynamic radius
may have a temperature dependence and ii) it is well known that in
supercooled liquid at low temperature the "decoupling" phenomenon
(the failure of the Stokes Einstein relation) occurs. In the
recent years, the fragility has been quantified according to the
$T$ behavior of $\eta$, but this has been done following different
prescription ({\it vide infra}).

Despite minor ambiguities introduced by its different definitions,
the concept of fragility has a deep influence on the study of
relaxation processes in supercooled liquids. Many studies have
evidenced the existence of correlations between the values of the
fragility and other properties of the supercooled liquids, such
as: i) the "visibility" of the Boson Peak \cite{Sokolov,Spyros};
ii) the $T$-dependence of the shear elastic modulus in liquids
(shoving model) \cite{Hall,Dyre1,Dyre2,Dyre3}; iii) the stretching
of the decay of the correlation functions at the glass transition
temperature \cite{BohmerAngell,Ngai}; iv) the nonlinearity of the
relaxation functions \cite{Hodge91} and, very recently, v) the
vibrational properties of the glass at $T \rightarrow 0$
\cite{Scopigno}. Other works have tried to extract physical
information on the nature of the glass transition from the
existence of these correlations \cite{Xia,Buchenau}. Finally, we
recall a recent attempt to extend the dimensionality of the space
spanned by the fragility index. Instead of using a single value
to classify the $T$-dependence of the viscosity, Ferrer et al.
\cite{Ferrer99} proposed to associate two indexes to every
glassformer. The first index (fragile/non-fragile) measures how
much the viscosity is Arrhenius- like at low temperature while
the second one (strong/weak) does the same around the melting
point. A deeper discussion on the correlation between fragility
and other supercooled liquid properties can be found in
Ref.~\cite{review}.

The relation between the statistical properties of the landscape
and the fragility is thought to be a central issue in the
comprehension of the physics behind the glass transition.
Debenedetti and Stillinger \cite{Debenedetti01} state in a very
recent review: "Equally important is the translation of
qualitative pictures ... into precise measures of strength and
fragility based on the basin enumeration function". A first
connection between the fragility and the topographic differences
in the energy landscape is found in Ref.~\cite{Stillinger95}.
There the landscapes of strong liquids were supposed to have a
"uniform" roughness, while a two-lengthscale arrangement of the
minima -with the introduction of the {\it meta-basins}, a concept
that has been recently revitalized by Doliwa and Heuer \cite{DH}-
was expected to characterize the PEL of fragile liquids. In 1995,
Angell \cite{Angell95}, rationalizing the much larger specific
heat jump at the glass transition shown by the fragile liquids
with respect to the strong ones, concluded that "Fragile liquids
would have high density of minima per unit energy..." and
"Surfaces with few minima ... generate strong liquids...". Similar
conclusions are reported in Ref.~\cite{Angell97} and by
Debenedetti and Stillinger \cite{Debenedetti01} who, more
recently, wrote that "... strong landscape may consist of a single
metabasin whereas fragile ones display a proliferation of
well-separated metabasin".

Summing up, there seems to be consensus on the statements
\begin{eqnarray}
\mbox{strong systems} &\Longleftrightarrow & \mbox{small} \ \
\alpha
\nonumber \\
\mbox{fragile system} &\Longleftrightarrow & \mbox{large} \ \
\alpha \nonumber
\end{eqnarray}

An attempt to determine a quantitative relation between fragility
and number of states on a theoretical basis, within the framework
of the "gaussian landscape model" (see below), is due to
Speedy\cite{Speedy99}, who derived a direct proportionality
between kinetic fragility and $\alpha$. This relation has been
then criticized by Sastry \cite{Sastry01}, who -again using the
gaussian model to fit his molecular dynamics simulation of the
Kob-Anderson Lennard Jones Binary Mixture (BMLJ) at different
densities- reached the conclusion of a proportionality between
fragility and the square root of $\alpha$.

In this paper we first present a summary of the different
definitions of "fragility" that are commonly used in the current
literature, and then recall several models of configurational
entropy (several "landscapes") proposed in the past that -with the
help of the Adam-Gibbs equation, or of the Vogel-Tamman-Fulcher
relation, or both- lead to different expression for the fragility
in terms of the parameters characterizing the "landscapes". In the
subsequent sections, we review  the Speedy and the Sastry
propositions on the $\alpha$- dependence of the fragility for the
examined landscapes. Finally we emphasize that landscapes with
the same statistical properties (i.e. same total number of
basins, same energy distribution of the basins depth) may be
characterized by different fragilities, calling attention on the
role of the different parameters entering in the Adam- Gibbs
expression. We conclude discussing the obtained results in the
context of the strong-to-fragile transition observed in some
strong glass forming liquids.

\section{Fragilities}

As discussed in the introduction, and following Angell
\cite{Angell85}, we will define the kinetic fragility in terms of
the temperature behavior of the viscosity and not of the
structural relaxation time. Having clarified this point,
however, we have to face -for the present purpose- different
definitions of the "index of (kinetic) fragility". The robustness
of a concept like the fragility lies in the observation that
-when plotting $\log(\eta(T))$ vs. $T/T_g$- the curves for
different liquids (beside very few exceptions) do not intersect
each other, and converge to a common point at $T=T_g$ (by
definition) and at $T\rightarrow\infty$. Given this situation, it
is possible to sort the systems, i.~e. to unambiguously asses
whether or not a system is more fragile than another. It is,
therefore, natural to assign a numerical value to this concept:
the index of fragility.

\subsection{Kinetic fragility: local definitions}

The first definition, let's call it "Angell's kinetic fragility",
$m_{_{A}}$, is
\begin{equation}
m_{_{A}}\doteq \frac{d \log(\eta(T)/\eta_\infty)}{d(T_g/T)} \Big
|_{T=T_g} \label{mA}
\end{equation}
Here $\eta_\infty$ is the limiting high temperature viscosity and
$T_g$ is defined from the condition $\eta(T_g)=10^{13}$ poise. As
it is experimentally observed that all the liquids share a very
similar value of $\eta_\infty \cong 10^{-4}$ poise, this quantity
is {\it conventionally} fixed to this value. Accordingly, an
ideal strong glass (strictly Arrhenius behavior) would have
$m_{_{A}} \cong $ 17, whereas higher values are indication of
higher fragility. While in principle there is no upper limit for
$m_{_{A}}$, on a practical ground the most fragile systems seems
to be tri-phenyl-phosphate, with $m_{_{A}}\approx 160$.

A very similar definition has been proposed by Speedy
\cite{Speedy99}:
\begin{equation}
m_{_{S}}\doteq \frac{d \left [
\frac{\log(\eta(T)/\eta_\infty)}{\log(\eta(T_g)/\eta_\infty)}\right
] }{d(T_g/T)} \Big |_{T=T_g} \label{mS}
\end{equation}
At a first sight, it seems that a trivial normalization factor
would bring from $m_{_{S}}$ to $m_{_{A}}$. However, this
expression become more useful than Eq.~1 if we want to relax the
{\it assumption} $\eta_\infty$ = $10^{-4}$ poise. In conjunction
with Eq.~\ref{mS}, it is also useful to define the glass
transition temperature $T_g$ as the temperature where
$\eta(T_g)/\eta_\infty=10^{17}$; we will use this definition
hereafter. As we will see below, if we aim to study, for example,
the density dependence of the fragility of a given system, it will
be easier to use Eq.~2 where the density dependence of
$\eta_\infty$, although small, has been washed out. It is worth to
point out, however, that for all the practical purposes, when
dealing with the experimental data the difference in using Eq.~1
or Eq.~2 is by all means irrelevant (apart from a trivial factor
very close to 17). The fragility index $m_{_{S}}$ ranges from one
for strong glasses to $\approx$10 for the more fragile systems.

The previous two definitions focus on the behavior of $\eta(T)$ at
the glass transition temperature. More recently, another index of
fragility ---often referred to as $F_{1/2}$--- has been introduced
by Richert and Angell \cite{Richert98} to "measure" the fragility
at intermediate temperature (see also the discussion in
Ref.~\cite{Green}). Naming $T^*$ the temperature that satisfies
$\log(\eta(T^*))=[\log(\eta(T_g))+\log(\eta(T_\infty))]/2$ (i.~e.
the temperature where the viscosity is halfway -in logarithmic
scale- between $\eta_\infty$ and 10$^{13}$ poise), $F_{1/2}$ is
defined as $F_{1/2}=2(T_g/T^*)-1$. It is worth to mention that
$F_{1/2}$ and $m_{_{A}}$ (or $m_{_{S}}$) do not provide "exactly"
the same information:  a plot of one quantity against the other
does not indicate a perfect correlation, rather it shows a scatter
of the points around an average trend \cite{Wang}. The existence
of such a scattering has been recently rationalized by Chandler
and Garrahan within the framework of a coarse-grained model of
glass formers \cite{Chandler03}.

Finally,  a generalized, temperature dependent fragility (either
$m_{_A}$ or $m_{_S}$) is sometime introduced, using equations
similar to Eq.s~\ref{mA} or \ref{mS} where $T_g$ is substituted by
a generic reference temperature $T$. We will call these quantities
as $m_{_A}(T)$ and $m_{_S}(T)$, with the implicit definition that
when the argument is missing, the quantities are calculated at
$T=T_g$.

\subsection{Kinetic fragility: global definitions}

The previous indexes of fragility were associated to the behavior
of $\eta(T)$ at a given temperature. Other definitions are based
on the global behavior of the viscosity, and necessarily rely on
the existence of a functional expression for $\eta(T)$.

A global definition of kinetic fragility arises from the
experimental observation that the temperature dependence of the
viscosity follows rather closely a Vogel-Tamman-Fulcher (VTF) law
\cite{VTF}:
\begin{equation} \eta(T)=\eta_\infty
\exp(\frac{DT_o}{T-T_o}), \label{VTF}
\end{equation}
where $\eta_\infty$,  $D$ and $T_o$ are system dependent
parameters. As long as the VTF description of $\eta(T)$ is
correct, one of the two parameters in the argument of the
exponential can be eliminated in favor of $T_g$ as -from the
definition of glass transition temperature- the following relation
holds~\cite{Angell91}:
\begin{equation}
T_g=T_o(1+\frac{D}{17\ \ln(10)}) \label{TgVTF}
\end{equation}
Plugging Eq.~\ref{VTF} in Eq.~\ref{mS}, and using Eq.~\ref{TgVTF},
one gets that the parameter $D$ is related to the previously
defined fragilities:
\begin{equation}
D=\frac{17 \
\ln(10)}{m_{_{S}}-1}
\label{D}
\end{equation}
and, therefore, can be assumed to be a further fragility index.
This index, which ranges from $\infty$ for strong liquids
(actually $D \approx 100$ for vitreous silica) to $\approx$~5 for
the fragile ones, is in same sense "weaker" than the other three
previously introduced, as its validity is based on the assumed
$T$-dependence of the viscosity (Eq.\ref{VTF}).

The assumption of the validity of the VTF law for the viscosity
also leads to a relation between the local fragility defined at
different temperatures. Indeed, recalling the definition of
$F_{1/2}$ and Eq.~\ref{mS}, one gets \cite{Richert98}
\begin{equation}
F_{1/2}=\frac{m_{_{S}}-1}{m_{_{S}}+1} \label{F}
\end{equation}


\subsection{Thermodynamic fragility}

An important step forward in relating the fragility with the PEL
properties has been certainly achieved with the introduction of
the "thermodynamic fragility" \cite{Martinez01}. Similarly to the
kinetic fragility which naturally emerges from the Angell plot
($\log(\eta)$ vs. $T_g/T$ for different systems), the vigor of the
concept of thermodynamic fragility arises from the temperature
dependence of the excess entropy $S_{_{ex}}(T)$, defined as the
difference between the entropy of the liquid and the entropy of
the stable crystal. On plotting $S_{_{ex}}(T_g)/S_{_{ex}}(T)$ vs.
$T_g/T$, one obtains a plot very similar to the Angell plot, where
the different systems stand in the same order \cite{nota2}.

In similar fashion to the kinetic fragility $F_{1/2}$, it has been
defined a "thermodynamic" fragility $F_{3/4}$: naming  $T^*$ the
temperature where $S_{_{ex}}(T_g)/S_{_{ex}}(T)=3/4$, i.~e. the
temperature where the inverse excess entropy equals 3/4 of its
$T_g$ value, $F_{3/4}$ is defined as $F_{3/4}=2(T_g/T^*)-1$. In
this case, the value 3/4, and not 1/2, has been chosen because of
the difficulties associated to determine the excess entropy at
high $T/T_g$ in strong liquids. In a recent paper Martinez and
Angell \cite{Martinez01} have shown that it exists a remarkable
correlation between $F_{1/2}$ and $F_{3/4}$: with few exceptions
it turns out that $F_{1/2} \approx F_{3/4}$ within 10\%. This
observation rationalizes the well known fact that the amplitude of
the specific heat jump at $T_g$ is linked to the fragility, but
also points out that is not the specific heat jump alone, but
rather this jump divided by the excess entropy at $T_g$, that is
actually related to $m_{_A}$.

In analogy with $m_{_{A}}$ (or with $m_{_{S}}$) it would be
natural to define a further index of the thermodynamic fragility
as the derivative at $T_g$ of the inverse reduced excess entropy
with respect to the inverse reduced temperature. To our knowledge,
this index has not been yet introduced, but -as we will see
below- this quantity naturally appears when the Adam-Gibbs
relation is used to work out a link between kinetic and
thermodynamic fragility. It is useful, therefore, to introduce
this thermodynamic fragility ($m_{_{T}}$) index as:
\begin{equation}
m_{_{T}}\doteq \frac{d (S_{_{ex}}(T_g)/S_{_{ex}}(T))}{d(T_g/T)}
|_{T=T_g} =T_g\frac{S'_{_{ex}}(T_g)}{S_{_{ex}}(T_g)}\label{mT},
\end{equation}
being $S'_{_{ex}}(T)$ the temperature derivative of
$S_{_{ex}}(T)$.

\subsection{Relation between \\ kinetic and thermodynamic fragility}

The Adam-Gibbs equation \cite{Adam65} establishes a relation
between the structural relaxation time and the configurational
entropy $\Sigma(T)$:
\begin{equation}
\tau(T)=\tau_\infty \exp(\frac{\cal{E}}{T \Sigma(T)}) \label{AGt}
\end{equation}
or, relying on the Maxwell relation, between the viscosity and the
configurational entropy:
\begin{equation}
\eta(T)=\eta_\infty \exp(\frac{\cal{E}}{T \Sigma(T)}) \label{AG}
\end{equation}
where $\tau_\infty$ ($\eta_\infty$) is the usual infinite
temperature limit for the relaxation time (viscosity) and
$\cal{E}$ a system dependent parameter with the physical dimension
of an energy that is somehow related to the energy barrier for
activated processes. This equation is the key relation that allows
us to create a link between kinetic and thermodynamic fragility
and, ultimately, via the configurational entropy a link between
kinetic fragility and the statistical properties of the PEL. Let
us first observe that, as the energy barrier is expected to have a
weak and smooth temperature behavior and not to diverge at any
temperature, according to Eq.~\ref{AG} the viscosity diverges at
the temperature (Kauzmann temperature $T_K$) where the
configurational entropy vanishes. If both the Adam-Gibbs
(Eq.~\ref{AG}) and Vogel-Tamman-Fulcher relations (Eq.~\ref{VTF})
are valid, then necessarily $T_o$ and $T_K$ are equal one to each
other. This equality has been recently disputed \cite{Tanaka03}.
We do not further discuss this problem, with the aim to study the
mathematical consequences of the different landscape models
introduced in the literature, we will assume (when necessary) that
$\cal E$ is a slowly varying smooth function of $T$ (thus, that
$T_o=T_K$). It must also be noted that the thermodynamic fragility
is defined through the experimentally accessible {\it excess}
entropy, while the Adam-Gibbs relation calls into play the {\it
configurational} entropy. In the following we will not make
difference between the two entropies, relying upon the observation
that configurational and excess entropy seems to be actually
proportional to each other \cite{Corezzi02}, even if other studies
indicate the failure of such a proportionality \cite{S1}. Assuming
that the Adam-Gibbs relation correctly describes the
$T$-dependence of the viscosity in a supercooled liquids, by
plugging Eq.~\ref{AG} into the definition of $m_{_{S}}$,
Eq.~\ref{mS}, we get (using $\eta(T_g)/\eta_\infty=10^{17}$):
\begin{equation} m_{_{S}} =
1+T_g\frac{\Sigma'(T_g)}{\Sigma(T_g)} \label{relST},
\end{equation}
and, recalling Eq.~\ref{mT}, we have the desired relation between
kinetic and thermodynamic fragility:
\begin{equation}
m_{_{S}} = 1 + m_{_{T}}.
\end{equation}
Equation \ref{relST} also constitutes the basis to obtain a link
between the kinetic fragility $m_{_{S}}$ and the number of states
$\alpha$. Indeed, recalling the relation
$\alpha=max_e\{\Sigma(e(T))\}/Nk_B$, if we know -or have a model
for- the configurational entropy of a given system, we could
determine $\alpha$ and $m_{_{S}}$, and thus try to relate one to
the other.

\section{Models of landscape}

In this section we will briefly recall the main models that have
been introduced in the recent literature to represent the
configurational entropy of supercooled liquid systems. In the
first three subsections we elucidate models of configurational
entropy and derive the relations between the different quantities
of interest ($T$ and $e$ dependence of $\Sigma$, fragility, etc.)
with the specific hypothesis that the vibrational entropy
associated to a specific minimum of the PEL is independent from
its energy elevation. In the following subsection, we relax this
hypothesis, assuming a linear dependence of the vibrational free
energy from $e$, and showing how the equations relating the
relevant physical quantities to the configurational entropy
parameters are modified.

\subsection{Gaussian model}

The gaussian model is at the basis of the interpretation of the
configurational entropy in simulated supercooled liquids. After
the first studies\cite{S3,S4,otp}, the gaussian model has been
chosen to describe quantitatively the  energy dependence of
$\Sigma(e)$ in different systems
\cite{Speedy99,ModGaus1,Sastry01,ModGaus3,ModGaus4}. According to
this model, an explicit functional form (gaussian) for
$\Omega_N(E)$ -the energy distribution of the minima of the PEL-
is assumed,
\begin{equation}
\Omega_N(E) = \exp({\alpha N}) \exp
[-\frac{(E-E_o)^2}{\epsilon^2}], \label{GM}
\end{equation}
From this equation, the configurational entropy of the gaussian
model becomes ($e=E/N$):
\begin{equation}
\Sigma(e) = k_B N \left [ \alpha - \frac{(e-e_o)^2}{\bar
\epsilon^2} \right ] \label{GM2}
\end{equation}
being $\bar \epsilon = \epsilon / \sqrt{N}$. In this expressions
$\alpha$ counts the total number of states (it is the maximum of
$\Sigma(e)/N$ in $k_B$ units), $e_o$ is an irrelevant parameter
(it fixes the zero of the energy scale) and $\bar \epsilon$ is
the width of the distribution. In order to express the
configurational entropy as a function of the temperature, we must
first determine the energy of the minima of the PEL populated at
a given temperature. Using \cite{nota3,simTV}
\begin{equation}
\frac{1}{T}=\frac{d\Sigma(e)/N}{d e} \label{MS}
\end{equation}
we get
\begin{equation}
e(T)=e_o - \frac{\bar\epsilon^2}{2k_BT } \label{et}
\end{equation}
and, finally, inserting Eq.~\ref{et} into Eq.~\ref{GM2}, we have
the explicit expression of the configurational entropy as a
function of the temperature:
\begin{equation}
\Sigma(T) = k_B N
\left [ \alpha - \frac{\bar \epsilon^2}{(2k_BT)^2} \right ]
\label{GM3}
\end{equation}

From Eq.~\ref{GM2}, the Kauzmann energy $e_K$, i.e. the energy
where $\Sigma(e)=0$, is promptly derived:
\begin{equation}
e_K = e_o - \bar \epsilon \sqrt{\alpha} \label{ek}
\end{equation}
and, plugging the Kauzmann energy (Eq.~\ref{ek}) in Eq.~\ref{et},
we find the Kauzmann temperature:
\begin{equation}
k_BT_K=\frac{\bar\epsilon}{2\sqrt{\alpha}}. \label{tk}
\end{equation}

It is useful to eliminate $\bar \epsilon$ from the expression of
the configurational entropy (in its explicit $T$-dependent
expression) in favor of $T_k$, using Eq.~\ref{tk}, so to obtain:
\begin{equation}
\Sigma(T) = k_B N \alpha \left [ 1 - \frac{T_k^2}{T^2} \right ]
\label{GM5}
\end{equation}

Once we have a model for the configurational entropy, we can
-applying Eq.~\ref{relST}- find an expression for the fragility in
terms of the parameters of the model itself. As parameters, we
have the freedom to choose among ($\alpha$, $\bar\epsilon$, $T_K$,
$e_K$). One compact possibility, which has the advantage to
explicitly depend only on $T_K$, is:
\begin{equation} m_{_{S}} =
\frac{T_g^2+T_K^2}{T_g^2-T_K^2}. \label{ms1}
\end{equation}
In this expression, $T_g$ appears explicitly and cannot be
eliminated because in the gaussian model (a {\it pure}
thermodynamic model) the dynamics is not defined and therefore
$T_g$ must be regarded as a parameter {\it external} to the
theory. Other possible selection of parameters, and thus other
expressions for the fragility, are of course possible.
Eq.~\ref{ms1} (as well as similar expressions for other landscape
models, see below) makes clear the well know fact that the
fragility is somehow related to the "distance" between $T_g$ and
$T_K$: the higher is the ratio $T_g/T_K$ the strongest is the
liquid.

As finally remark, we observe how -having imposed the validity of
both the Adam-Gibbs relation and the gaussian model for the
configurational entropy- the temperature dependence of the
viscosity turns out to be controlled by the law:
\begin{equation} \eta(T)=\eta_\infty
\exp(\frac{D T_K}{T-T_K}\frac{T}{T+T_K}), \label{VTF3}
\end{equation}
with $D ={\cal{E}}/(\alpha N k_B T_K)$, which is different by a
VTF relation. In other words, the VTF law, the Adam-Gibbs relation
and the gaussian model cannot be simultaneously invoked
(especially when the shape of the PEL basins is independent on the
depth). Equation~\ref{VTF3} can be regarded as a VTF law with a
{\it temperature dependent} coefficient $D'(T)=D T/(T+T_k)$. In
the high $T$ limit ($T>>T_k$), $D' \rightarrow D$ while in the low
$T$ regime ($T$ approaching $T_k$) $D' \rightarrow D/2$.

\subsection{Hyperbolic model}
For thirty years it has been realized \cite{Sichina} that the
temperature dependence of the (constant volume) excess specific
heat can be described by a hyperbolic law ($C\approx const +
const'/T$), and this law is commonly used to represent the
experimental data \cite{Richert98}. The "landscape model" that
gives rise to such a temperature dependence for the excess
specific heat is the so called hyperbolic model, recently
introduced and discussed in detail by Debenedetti, Stillinger and
Lewis \cite{DebStiLew}. In Ref.~\cite{DebStiLew}, the model is
derived from the assumption of a hyperbolic temperature dependence
of the "configurational" heat capacity, and (assuming the validity
of the Adam-Gibbs relation), it implies as a mathematical
consequence the validity of the VTF relation. For simplicity, here
we prefer to start assuming the mathematical validity of both the
Adam-Gibbs and the VTF, the hyperbolic temperature dependence of
the excess specific heat results as consequence. Obviously, as
discussed in \cite{DebBook} the two routes are equivalent. It is
worth to point out that the "gaussian landscape" is named after
the $e$-dependence of the number of states, while the "hyperbolic
landscape" is named after the $T$ behaviour of the specific heat
\cite{Richert98}, a rather different quantity. It is our aim to
write down the main expressions for this model using the same
notation of the previous section, and to extract the equations for
the fragilities. By comparing Eq.~\ref{VTF} and \ref{AG}, it turns
out an explicit temperature dependence for $\Sigma(T)$:
\begin{equation}
\Sigma(T) = \frac{\cal{E}}{D T_K}\left [ 1 - \frac{T_K}{T} \right
] \label{VTFM1}
\end{equation}
It is implicit in this expression the coincidence of $T_o$ and
$T_K$. This equation can be cast in form very similar to
Eq.~\ref{GM3} by defining the quantities $\alpha$ and $\bar
\epsilon$:
\begin{equation}
\alpha \doteq \frac{\cal{E}}{D N k_B T_K}, \label{al}
\end{equation}
\begin{equation}
\bar \epsilon \doteq \frac{2\cal{E}}{D N}=2 k_B T_K \alpha,
\label{tk2}
\end{equation}
As we will see soon, $\alpha$ and $\bar \epsilon$ play here the
same role as they have in the gaussian model, therefore the first
equation is a link between the "number of states" and the
constants entering in the AG ($\cal{E}$) and VTF ($D$ and $T_K$)
relations. The second equation can be compared to Eq.~\ref{tk},
where $\sqrt{\alpha}$ appears instead of $\alpha$. Rewriting
Eq.~\ref{VTFM1} with the elimination of $\cal{E}$ and $D$ in
favor of $\alpha$ and $\bar\epsilon$, we have:
\begin{equation}
\Sigma(T) = k_B N \left [ \alpha - \frac{\bar \epsilon}{2 k_B T}
\right ], \label{VTFM2}
\end{equation}
an expression that can be directly compared with Eq.~\ref{GM3},
or, expressing the pre-factor in Eq.~\ref{VTFM1} in terms of
$\alpha$  via Eq.~\ref{al},
\begin{equation}
\Sigma(T) = k_B N \alpha \left [ 1 - \frac{T_K}{T} \right ]
\label{VTFM3}
\end{equation}
that can be compared with Eq.~\ref{GM5}

At variance with the gaussian model, where we started with a model
for $\Sigma(e)$ and derived $\Sigma(T)$, we now have a model for
$\Sigma(T)$. To obtain an expression for $\Sigma(e)$ we first
derive the temperature dependence of the energy of the minima
visited by Eq.~\ref{MS}:
\begin{eqnarray}
\label{et2}
e(T)&=&e_R + \int_{T_R}^T T \ \frac{d\Sigma}{dT} \ dT \\
&=&e_R + \alpha k_B T_K \ln{(T/T_R)}  \nonumber
\end{eqnarray}
where $e_R$ and $T_R$ are integration constants whose values, as
we will see, are not relevant for the interesting physical
quantities. Inverting Eq.~\ref{et2} and plugging the resulting
$T(e)$ into Eq.~\ref{VTFM2} we get:
\begin{equation}
\Sigma(e) = N k_B \alpha \left [ 1 - \frac{T_K}{T_R} \exp{\left (
-\frac{2(e-e_R)}{\bar\epsilon}\right )} \right ] \label{VTFM4}
\end{equation}
Obviously, we can eliminate $T_R$ from this equation, by properly
redefining $e_R$. A useful possibility  is to choose $T_R=T_K$,
then, from Eq.~\ref{et2}, $e_R=e_K$ and:
\begin{equation}
\Sigma(e) = N k_B \alpha \left [ 1 - \exp{\left (
-\frac{2(e-e_K)}{\bar\epsilon}\right )} \right ]. \label{VTFM5}
\end{equation}
At variance with the configurational entropy of the gaussian
model, the present $\Sigma(e)$ does not show any maxima, rather it
increase continuously, asymptotically approaching the value
$Nk_B\alpha$.

From Eq.~\ref{VTFM3}, using Eq.~\ref{relST}, we can easily
determine the fragility of this model:
\begin{equation}
m_{_{S}} = \frac{T_g}{T_g-T_K}. \label{ms2}
\end{equation}
It is worth to point out that this expression is the expansion of
the fragility of the gaussian model to first order in $T_g-T_K$.

\subsection{Logarithmic {\bf (or binomial)} model}

The previous two models for the configurational entropy share the
property that $d\Sigma/de$ is non diverging at $e=e_K$, so the
Kauzmann temperature exists and it is non-vanishing. In order to
introduce a more flexible model, embedding the possibility of
having a vanishing Kauzmann temperature, Debenedetti, Stillinger
and Shell \cite{DebStiShell} recently proposed a modification of
the gaussian model that, with a slight change in notation with
respect to the original definition, reads:
\begin{eqnarray}
\label{LM1}
&& \Sigma(e) = N k_B \alpha \Big \{ (1-\gamma) \Big[1 -
\Big(\frac{u}{\sqrt{\alpha}}\Big)^2 \Big]+ \\ \nonumber &&\gamma \Big [
1-\frac{(1+\frac{u}{\sqrt{\alpha}})\ln(1+\frac{u}{\sqrt{\alpha}})
+(1-\frac{u}{\sqrt{\alpha}})\ln(1-\frac{u}{\sqrt{\alpha}})}{2 \ln(2)}
\Big ] \Big \},
\end{eqnarray}
with $u$ ($-\sqrt{\alpha}<u<\sqrt{\alpha}$) given by:
\begin{equation}
u=\frac{e-e_o}{\bar \epsilon} \label{defu}
\end{equation}
This is a linear combination -weighted by the parameter $\gamma$-
of the parabolic configurational entropy typical of the gaussian
model and a term that depends on the logarithm of the energy. Here
we want to describe in detail the properties of this model for the
specific case $\gamma=1$, i.~.e. of a model that is totally
"logarithmic". The logarithmic model is essentially a binomial
distribution, i.e. model for  the thermodynamics of a gas of
binary excitations \cite{austenbm}. It has been used to model the
thermodynamics of supercooled liquids and the $T$-dependence of
the inherent structure energy \cite{austenbm}. Obviously, the
logarithmic term in Eq.~\ref{LM1} become dominant in the low-
$T$/low-$(e-e_K)$ region, therefore the model discussed in this
section can be thought as an approximation of the Debenedetti,
Stillinger and Shell model valid in the low $T$ limit. It is,
however, interesting to study such a model in the whole energy
range. Indeed, as we will see below, a visual inspection of the
function $\Sigma(e)$ indicates that this model and the gaussian
model represent very similar "landscapes", i.~e. very similar
distribution of the minima energy. Thus, we define the
"logarithmic" landscape as:
\begin{eqnarray}
\label{LM2}
&& \Sigma(e) \!=\! N k_B \alpha \times \\ \nonumber && \left [
1\!-\!\frac{(1\!+\!\frac{u}{\sqrt{\alpha}})\ln(1\!
+\!\frac{u}{\sqrt{\alpha}})\!+\!(1\!-\!\frac{u}{\sqrt{\alpha}})
\ln(1\!-\!\frac{u}{\sqrt{\alpha}})}{2\ln(2)}  \right ]
\end{eqnarray}
This expression for the configurational entropy has the properties
to vanish at $u=\pm \sqrt{\alpha}$, therefore the Kauzmann energy
results to be at $u=-\sqrt{\alpha}$ or explicitly
$e_K=e_o-\bar\epsilon \sqrt{\alpha}$. At this energy, the
derivative of $\Sigma(e)$ shows a logarithmic divergence, thus
implying that the Kauzmann temperature must vanish. Similarly to
the gaussian model, the parameter $e_o$ is the energy of the "top
of the landscape" and $\alpha$ represents the maximum of
$\Sigma(e)/Nk_B$. Using Eq.~\ref{defu} and the expression for
$e_K$, Eq.~\ref{LM2} can be explicitly written in terms of the
reduced energy measured with respect to the Kauzmann energy
$v=(e-e_K)/\bar\epsilon$ as:
\begin{eqnarray}
\Sigma(e) \!=\! N k_B \alpha \left [
1\!-\!\frac{\frac{v}{\sqrt{\alpha}}\ln(\frac{v}{\sqrt{\alpha}})
\!+\!(2\!-\!\frac{v}{\sqrt{\alpha}})
\ln(2\!-\!\frac{v}{\sqrt{\alpha}})}{2\ln(2)}  \right ] \label{LM2b}
\end{eqnarray}
We can now follow the same route used in the discussion of the
gaussian model. Via Eq.~\ref{MS}, with straightforward algebra, we
obtain the temperature dependence of the energy of the minima:
\begin{equation}
e(T)-e_o= -\bar\epsilon \sqrt{\alpha} \tanh \left [ \frac{2
\ln(2) \bar \epsilon}{\sqrt{\alpha} k_B T} \right ] \label{TdiE}
\end{equation}
and inserting this expression in Eq.~\ref{LM2b} the temperature
dependence of the configurational entropy is promptly derived:
\begin{eqnarray}
\Sigma(T) \!=\! N k_B \alpha &\Big \{& \frac{1}{\ln(2)} \ln \left
[ 2 \cosh \left ( \frac{\ln(2) \bar \epsilon}{\sqrt{\alpha} k_B T}
\right ) \right ] \nonumber \\ &-& \frac{\bar
\epsilon}{\sqrt{\alpha} k_B T}\tanh \left ( \frac{\ln(2) \bar
\epsilon}{\sqrt{\alpha} k_B T} \right ) \Big \} \label{LM3}
\end{eqnarray}

As a consequence of the infinite value of $\Sigma(e)/de$ at $e_K$,
this function does vanish only at $T=0$, i.~e. for this model
$T_K=0$. It is convenient, for sake of compactness, to define a
typical temperature which -in analogy with $T_K$ in the gaussian
and hyperbolic models- could be used to scale the temperatures in
the logarithmic model. We arbitrarily introduce the quantity:
\begin{equation}
T_K^*= \frac{1}{3} \ln(2) \frac{\bar\epsilon}{ k_B \sqrt{\alpha}}
\label{TkLM}
\end{equation}
whose value is very close to the "apparent" Kauzmann temperature
that would have been identified by extrapolating Eq.~\ref{LM3}
towards zero using only information on $\Sigma(T)$ at "high"
temperature, similarly to what is done experimentally. In other
words, the logarithmic model predicts a temperature dependence of
the configurational entropy that -around the inflection region-
can be approximated by a straight line that goes to zero at $k_B
T_K \sqrt{\alpha}/ \bar\epsilon \approx 0.23$ ($\approx
\ln(2)/3$). Having introduced the "apparent" Kauzmann temperature
for the logarithmic model, we can write Eq.s~\ref{TdiE} and
\ref{LM3} as:
\begin{equation}
e(T)-e_o= \frac{3}{\ln(2)} \alpha k_B T_K^*  \tanh \left ( \frac{3
T_K^*}{T} \right ) \label{TdiE2}
\end{equation}
\begin{eqnarray}
\Sigma(T) \!=\! \frac{N k_B \alpha}{\ln(2)} \left \{  \ln \left [
2 \cosh \left ( \frac{3 T_K^*}{T} \right ) \right ] \nonumber -
\frac{3 T_K^*}{T} \tanh \left( \frac{3 T_K^*}{T}\right ) \right \}
\label{LM4}
\end{eqnarray}

\begin{table*}[t]
\begin{tabular}{@{\hspace{5pt}} c @{\hspace{5pt}} |*{3}{@{\hspace{15pt}} c}}
\hline & & & \\& gaussian model &  hyperbolic model  &
logarithmic model   \\  & & &   \\
\hline

& & & \\

& \fbox{$ \displaystyle \alpha - u^2 $} & $\alpha \left [ 1 -
\exp{\left ( -2u \right ) } \right ]$ & \fbox{$ \displaystyle
\alpha \! \left [ \!1\!-\! \frac{1}{2\ln(2)}\left [
(1\!+\!\!\frac{u}{\sqrt{\alpha}})
\ln(1\!+\!\!\frac{u}{\sqrt{\alpha}})\!+\!(1\!-\!\!\frac{u}{\sqrt{\alpha}})
\ln(1\!-\!\!\frac{u}{\sqrt{\alpha}}) \right ] \right ] $}
\\

$\Sigma(e)/Nk_B \{$ & & & \\

& $ -v^2 +2 \sqrt{\alpha} v $ & $\alpha \left [ 1 - \exp{\left (
-2v \right ) } \right ]$ & $\alpha \left [
1\!-\!\frac{1}{2\ln(2)}\left [
\frac{v}{\sqrt{\alpha}}\ln(\frac{v}{\sqrt{\alpha}})
\!+\!(2\!-\!\frac{v}{\sqrt{\alpha}})
\ln(2\!-\!\frac{v}{\sqrt{\alpha}})  \right ]  \right ]  $
\\

& & & \\

$e_K$ & $e_o-\bar \epsilon \sqrt{\alpha}$ & $e_o$ & $e_o-\bar
\epsilon \sqrt{\alpha}$ \\

& & & \\

$k_B T_K$ & $\frac{\bar\epsilon}{2\sqrt{\alpha}}$ &
$\frac{\bar\epsilon}{2\alpha}$  & $\frac{\ln(2)}{3}
\frac{\bar\epsilon}{\sqrt{\alpha}}$ \\

& & & \\

$\bar\epsilon$ & ${2\sqrt{\alpha}}k_BT_K$ & ${2\alpha}k_BT_K$ & $
\frac{3}{\ln(2)} \sqrt{\alpha} k_B T_K^*$ \\

& & & \\

$e(T)-e_o$ & $-2\alpha k_B T_K \left( \frac{T_K}{T}\right )$ &
$\alpha k_B T_K \ln{\left ( \frac{T}{T_K} \right )}$  &
-$\frac{3}{\ln(2)} \alpha k_B T_K^*  \tanh \left ( \frac{3
T_K^*}{T} \right )$
\\

& & & \\

$e(T)-e_K$ & $2\alpha k_B T_K \left(1- \frac{T_K}{T}\right )$ &
$\alpha k_B T_K \ln{\left ( \frac{T}{T_K} \right )}$  &
$\frac{3}{\ln(2)} \alpha k_B T_K^*  \left [ 1-\tanh \left (
\frac{3 T_K^*}{T} \right )\right ]$
\\

& & & \\

$\Sigma(T)/Nk_B$ & $\alpha \left [ 1 - \left(\frac{T_K}{T}
\right)^2 \right ]$ & $  \alpha \left [ 1 - \left( \frac{T_K}{T}
\right ) \right ]$ & $\frac{\alpha}{\ln(2)} \left \{  \ln \left [
2 \cosh \left ( \frac{3 T_K^*}{T} \right ) \right ] \nonumber -
\frac{3 T_K^*}{T} \tanh \left( \frac{3 T_K^*}{T}\right ) \right
\}$
\\

& & & \\

$m_{_S}$ & $\frac{T_g^2+T_K^2}{T_g^2-T_K^2}$ &
$\frac{T_g}{T_g-T_K}$ & $1\!+\!w^2\! \Big\{ \cosh^2(w)\ln \left
[2\cosh(w) \right]\! - w\cosh(w)\sinh(w) \Big\}^{-1} $
\\

& & & \\

$\ln{\left( \eta(T)/\eta_\infty \right )}$ & ${\frac{D
T_K}{T-T_K}\frac{T}{T+T_K} }$ & \fbox{$ \displaystyle{ \frac{D
T_K}{T-T_K} }$} & $ \Big \{ \frac{\ln(2)}{3} D \frac{3T_K^*}{T}
\Big \{  \ln \left [ 2 \cosh \left ( \frac{3 T_K^*}{T} \right )
\right ]- \frac{3 T_K^*}{T} \tanh \left( \frac{3 T_K^*}{T}\right )
\Big \}^{-1} \Big \}$
\\

& & & \\
\hline
\end{tabular}
\caption{Summary of the main relations relating the relevant
quantities (left column) for the three configurational entropy
models introduced before. The relation that define the model is
reported in box. The variable $u$ is defined as the reduced energy
measured with respect to $e_o$: $u=(e-e_o)/\bar\epsilon$, while
$v$ is that measured starting from the Kauzmann energy:
$v=(e-e_K)/\bar\epsilon$. The variable $w$ is a shortcut for
$3T_K^*/T_g$. In the case of the logarithmic model, $T_K^*$ is
used (see Eq.~\ref{TkLM}) in place of the Kauzmann temperature.}
\end{table*}

\begin{figure}[ht]
\centering
\includegraphics[width=.43\textwidth]{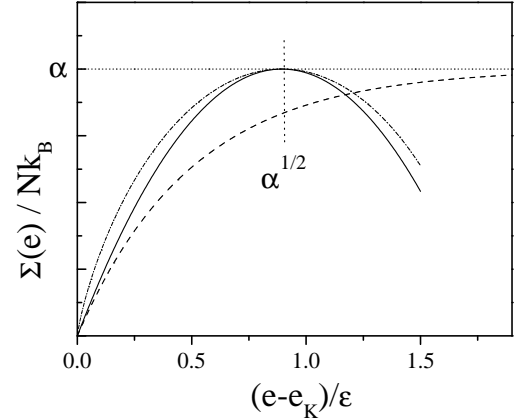}
\vspace{-4.7cm} \caption{Sketch of the energy dependence of the
configurational energy for three models: gaussian (full line),
hyperbolic (dashed line) and logarithmic (dot-dashed line). The
reduced entropy $\Sigma(e)/Nk_B$ is plotted as a function of
$(e-e_K)/\bar\epsilon$ for the specific case of $\alpha=0.8$}
\label{f1}
\end{figure}

Once the explicit $T$ dependence of $\Sigma(T)$ is known, both the
fragility $m_{_S}$, definite in Eq.~\ref{relST}, and the $T$
dependence of the viscosity (from the Adam-Gibbs equation) can be
worked out. The two expressions reads:
\begin{eqnarray}
m_{_S} &\!=\!&1\!+\!\left(\frac{3T_K^*}{T_g}\right)^2\! \Big\{
\cosh^2\!\!\left(\frac{3T_K^*}{T_g}\right)\ln \left
[2\cosh\!\left(\frac{3T_K^*}{T_g}\right)\right]\! \nonumber \\
&-& \left(\frac{3T_K^*}{T_g}\right)\cosh\!
\left(\frac{3T_K^*}{T_g} \right)\sinh \!
\left(\frac{3T_K^*}{T_g}\right) \Big\}^{-1} \label{ms3}
\end{eqnarray}
\begin{eqnarray}
\eta(T)&=& \eta_\infty \exp \Big \{ \frac{\ln(2)}{3} D
\left(\frac{3T_K^*}{T_g}\right) \Big \{  \ln \left [ 2 \cosh
\left ( \frac{3 T_K^*}{T} \right ) \right ]  \nonumber \\ &-&
\left(\frac{3T_K^*}{T_g}\right) \tanh \left( \frac{3
T_K^*}{T}\right ) \Big \}^{-1} \Big \}. \label{etalog}
\end{eqnarray}

\subsection{Summary of models}

In Table I, we summarize the expressions derived in the
framework of the three model examined before for different
quantities. These quantities are:
\begin{description}
\item{i) and ii)} the configurational entropy as a
function of $e$, in this case we explicitly report $\Sigma(e)$ as
a function of the variables $u=(e-e_o)/\bar\epsilon$ and
$v=(e-e_K)/\bar\epsilon$ to emphasize that the zero of the energy
is irrelevant and that $\bar\epsilon$ only acts as an energy
scale.
\item{iii)} The explicit expression of the Kauzmann energy in
term of $e_o$, $\alpha$ and $\bar\epsilon$.
\item
 iv) and v) The
relations used to eliminate $\bar\epsilon$ in favor of $T_K$ (or
$T_K^*$ in the case of the logarithmic model).
\item vi) and vii) The
temperature dependence of the inherent structures energy,
reported in terms of $T_K$.
\item viii) The temperature dependence of
the configurational entropy, now reported in term of $T_K$. \item ix)
The expression for the fragility reported in terms of the thermal
parameters.
\item x) Finally, we report the temperature dependence of
the viscosity resulting from the application of the model. In the
last expression the parameter $D$ is: $D={\cal{E}}/(\alpha N k_B T_K)$.
\end{description}

\begin{figure}[ht]
\centering
\includegraphics[width=.43\textwidth]{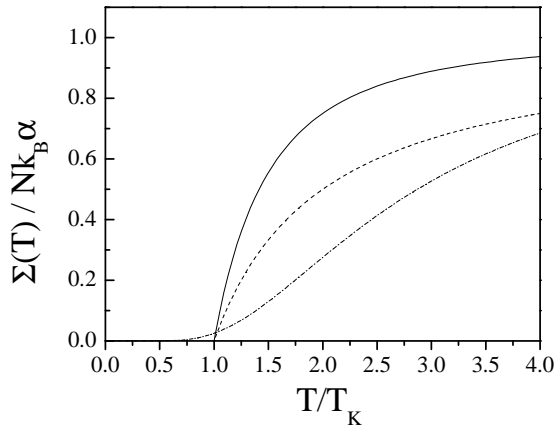}
\vspace{-5.cm} \caption{Plot of the temperature dependence of the
configurational entropy for three models: gaussian (full line),
hyperbolic (dashed line) and logarithmic (dot-dashed line). The
reduced entropy $\Sigma(T)/Nk_B$ further normalized to $\alpha$ is
plotted as a function of $T/T_K$. In the case of the logarithmic
model $T_K^*$ defined in Eq.~\ref{TkLM} is used in substitution of
$T_K$.} \label{f2}
\end{figure}

\begin{figure}[ht]
\centering
\includegraphics[width=.43\textwidth]{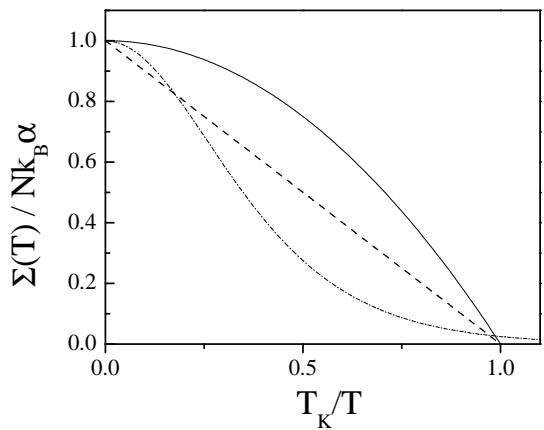}
\vspace{-5.1cm} \caption{Similarly to Fig.~\ref{f2}, the
temperature dependence of the configurational entropy for three
models (gaussian (full line), hyperbolic (dashed line) and
logarithmic (dot-dashed line)) is plotted as a function of
$T_K/T$. In the case of the logarithmic model $T_K^*$ defined in
Eq.~\ref{TkLM} is used in substitution of $T_K$.} \label{f3}
\end{figure}

In Fig.~\ref{f1} we sketched the $e$ dependence of the
configurational energy for the examined models: gaussian (full
line), hyperbolic (dashed line) and logarithmic (dot-dashed line).
As an example, the three configurational entropies are reported
for the specific case of $\alpha=0.8$ (as the scaling of
$\Sigma(e)$ with $\alpha$  for the hyperbolic model is different
from that for the gaussian and logarithmic models, we cannot use
a reduced variable).

Similarly, in Fig.s~\ref{f2} and \ref{f3} we report the
corresponding configurational entropy as a function of $T/T_K$
and $T_K/T$ respectively. In the case of the logarithmic model
$T_K^*$ defined in Eq.~\ref{TkLM} is used to scale the
temperatures.

\begin{figure}[ht]
\centering
\includegraphics[width=.43\textwidth]{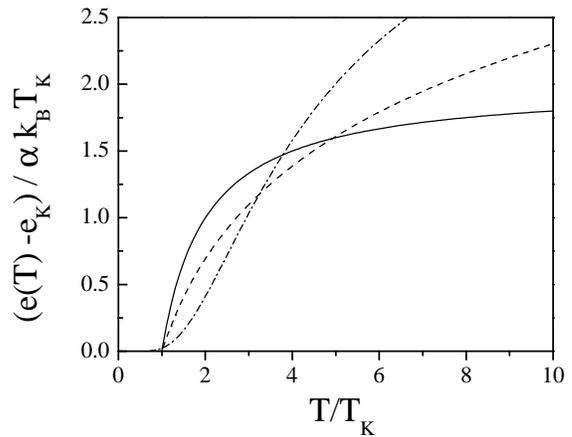}
\vspace{-5.1cm} \caption{Plot of the temperature dependence of the
energy elevation for three models: gaussian (full line),
hyperbolic (dashed line) and logarithmic (dot-dashed line). The
energy elevation $e-e_K$, normalized to $k_B T_K$, and further
normalized to $\alpha$, is plotted as a function of $T/T_K$. In
the case of the logarithmic model $T_K^*$ defined in
Eq.~\ref{TkLM} is used in substitution of $T_K$.} \label{f4}
\end{figure}

In Fig.~\ref{f4} we report the temperature dependence of the
energy elevation (normalized to the factor $\alpha k_B T_K$) with
respect to $e_K$ of the minima of the PEL visited at equilibrium
for the three examined models: gaussian (full line), hyperbolic
(dashed line) and logarithmic (dot-dashed line). The hyperbolic
model shows a non physical continued rise of $e(T)$ at increasing
$T$.

\begin{figure}[ht]
\centering
\includegraphics[width=.43\textwidth]{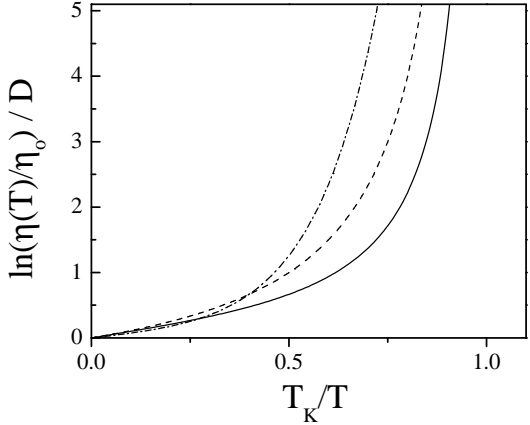}
\vspace{-5.1cm} \caption{Plot of the temperature dependence of the
viscosity for three models: gaussian (full line), hyperbolic
(dashed line) and logarithmic(dot-dashed line). The logarithm of
the viscosity normalized by $\eta(T_\infty)$, normalized to $D$
is plotted as a function of $T_K/T$. In the case of the
logarithmic model $T_K^*$ defined in Eq.~\ref{TkLM} is used in
substitution of $T_K$.} \label{f5}
\end{figure}

Finally, in Fig.~\ref{f5} we report in Arrhenius scale the
temperature dependence of the viscosity for the three examined
models: gaussian (full line), hyperbolic (dashed line) and
logarithmic (dot-dashed line).

\subsection{Gaussian models with \\ non-constant vibrational entropy}

All the discussion in the previous sections was based on the
assumption that the {\it vibrational} entropy associated to a
given basin is independent from the energy elevation of the
minimum of the basin itself. These assumption lead to the
simplified microcanonical definition of temperature reported in
Eq.~\ref{MS}. Following recent experimental \cite{S1} and
numerical \cite{Sastry01,ModGaus3} evidences indicating a
vibrational entropy that actually depends on the energy of the
minima, in the present section we relax the previous assumption,
and, for the specific case of the gaussian model, we develop the
calculation in the case of an explicit dependence of the
vibrational entropy, $S_v$, on $e$. In particular, taking
advantage of the outcome of recent molecular dynamics
calculations, we develop $S_v(e)$ in series of $e-e_K$ and retain
only the first order term, an approximation certainly valid for
low enough temperature:
\begin{equation}
S_v(e) = S_v^K + \frac{dS_v}{de} \Big \vert_{e=e_K} (e-e_K)
\label{Svib}
\end{equation}
The quantity $dS_v/de$ is a further system-dependent parameter.
For sake of simplicity let us define as parameter a "vibrational"
temperature $T_v$ via:
\begin{equation}
\frac{N}{T_v} = \frac{dS_v}{de} \Big \vert_{e=e_K}. \label{Tvib}
\end{equation}
The calculation proceeds along the same line outlined in the case
of the gaussian model. First, from the generalization of
Eq.~\ref{MS}, i.~e. from:
\begin{equation}
\frac{1}{T}=\frac{dS/N}{d e}= \frac{d\Sigma(e)/N}{d e}+
\frac{dS_v/N}{de}=\frac{d\Sigma(e)/N}{d e}+ \frac{1}{T_v}
\label{MS2}
\end{equation}
we get the temperature dependence of the energy of the visited
minima:
\begin{equation}
e(T)=e_o - \frac{\bar\epsilon^2}{2} \left(
\frac{1}{k_BT}-\frac{1}{k_BT_v} \right )  \label{et2b}
\end{equation}
and inserting Eq.~\ref{et2b} into the definition of the gaussian
model, Eq.~\ref{GM2}, we have the explicit expression of the
configurational entropy as a function of the temperature:
\begin{equation}
\Sigma(T) = k_B N \left [ \alpha - \frac{\bar \epsilon^2}{4}
\left( \frac{1}{k_BT}-\frac{1}{k_BT_v} \right )^2 \right ]
\label{GM13}
\end{equation}
We can now eliminate $\bar \epsilon$ by introducing the Kauzmann
temperature defined by $\Sigma(T_K)=0$:
\begin{equation}
\bar\epsilon= 2\sqrt{\alpha} k_B T_K \left( \frac{T_v}{T_v-T_K}
\right )
\end{equation}
thus, substituting this expression in Eq.~\ref{GM13}
\begin{equation}
\Sigma(T) = k_B N \alpha \left [ 1 - \left (\frac{T_K}{T_v-T_K}
\right)^2 \left ( \frac{T_v}{T} -1 \right )^2 \right ]
\label{GM5b}
\end{equation}

Through the configurational entropy, we can apply Eq.~\ref{relST}
to find an expression for the fragility:
\begin{equation} m_{_{S}} =
\frac{(T_g^2+T_K^2)-2T_gT_K (T_g/T_v)
}{(T_g^2-T_K^2)-2T_K(T_g-T_K)(T_g/T_v)}. \label{ms1b}
\end{equation}
In this expression, besides $T_g$ -the parameter that embodies our
choice of the value of viscosity that define the glass transition
temperature- there are the two system-dependent parameter $T_K$
(a way to express $\bar \epsilon$) and $T_v$. Finally the
temperature dependence of the viscosity turns out to be
controlled by the law:
\begin{eqnarray}
\eta(T)=\eta_\infty \exp \Big (  &&\frac{D
T_K}{T-T_K}\frac{T}{T+T_K-2T(T_K/T_v)} \nonumber \\ && \left [
\frac{T_v-T_K}{T_v} \right]^2  \Big ), \label{VTF3b}
\end{eqnarray}
with, as before, $D ={\cal{E}}/(\alpha N k_B T_K)$.

\begin{figure}[ht]
\centering
\includegraphics[width=.43\textwidth]{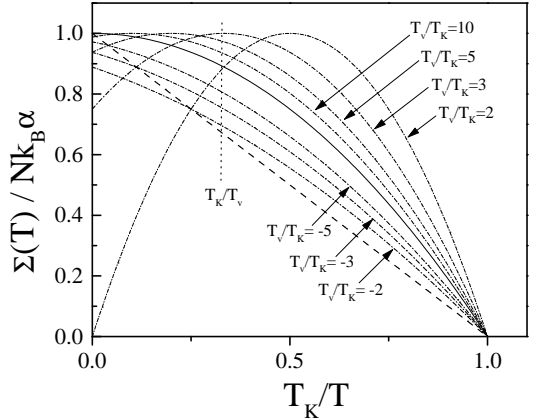}
\vspace{-5.cm} \caption{Similar to Fig.~\ref{f2}, the temperature
dependence of the configurational entropy is reported for the
gaussian model with energy dependent vibrational entropy for
different values of the parameter $T_v/T_K$ (dot-dashed line). For
comparison, also the two corresponding function for the gaussian
(full line) and hyperbolic (dashed line) models are reported.}
\label{f6}
\end{figure}

Similarly to Fig.~\ref{f3}, in Fig.~\ref{f6} we report the
temperature dependence of the configurational entropy of the
gaussian model with energy dependent vibrational entropy as a
function of $T_K/T$  for different values of $T_v/T_K$ (reported
in the figure) and compared with the similar quantity for the
gaussian and the hyperbolic models.

\begin{figure}[h]
\centering
\includegraphics[width=.43\textwidth]{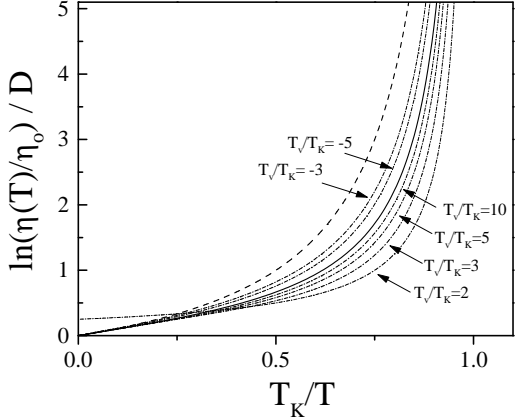}
\vspace{-5.1cm} \caption{Similar to Fig.~\ref{f5}, the temperature
dependence of the viscosity is reported for the gaussian model
with energy dependent vibrational entropy for different values of
the parameter $T_v/T_K$  (dot-dashed line). For comparison, also
the two corresponding function for the gaussian (full line) and
hyperbolic (dashed line) models are reported.} \label{f7}
\end{figure}

Analogously, Fig.~\ref{f7} shows the temperature dependence of the
viscosity as predicted by the gaussian model with energy dependent
vibrational entropy for different values of $T_v/T_K$. As can be
noticed, it seems that the values of $T_v$ allows one to
interpolate between the behavior of the gaussian model (obviously
reached for $T_v \rightarrow \infty$ or $-\infty$) and that of the
hyperbolic model (that is approximately obtained for $T_v/T_K
\approx -1 \div -1.5$). It is worth to remember that, in most
numerical simulations of model liquids, $T_v$ is found to be
negative for constant density (thus constant PEL) simulations,
while $T_v > 0$ for constant pressure simulations \cite{ConstP}.
In the case of a model for water, the sign of $T_v$ has been found
to be density dependent \cite{ModGaus4,newprl}. On the
experimental side, at constant pressure, the sign of $T_v$ turns
out to be both positive \cite{S1} and negative \cite{S2},
depending on the specific system.

Finally in Table II we report the relevant expression relative to
the gaussian model with energy dependent vibrational entropy
($T_v\neq \infty$) compared with those of the gaussian model
($T_v=\infty$).

\begin{table*}[t]
\begin{tabular}{@{\hspace{5pt}} c @{\hspace{5pt}} |*{2}{@{\hspace{20pt}} c}}
\hline & &  \\& $\frac{dS_v}{de}=0$ &   $\frac{dS_v}{de} \neq 0$  \\
& &
  \\ \hline

%
%
%

& &  \\

$k_B T_K$ & $\frac{\bar\epsilon}{2\sqrt{\alpha}}$ & $\frac{k_B T_v
\bar\epsilon}{2\sqrt{\alpha}k_B T_v+\bar\epsilon}$ \\

& &  \\

$\bar\epsilon$ & ${2\sqrt{\alpha}}k_BT_K$ &
${2\sqrt{\alpha}}k_BT_K \left [ \frac{T_v}{T_v-T_K} \right ]$ \\

& &  \\

$e(T)-e_o$ & $2\alpha k_B T_K \left( \frac{T_K}{T}\right )$ &
${2{\alpha}}k_BT_K \left [ \frac{T_v T_K}{(T_v-T_K)^2} \right ]
\left [ \frac{T_v}{T} -1 \right ]$\\

& &  \\

$\Sigma(T)/Nk_B$ & $\alpha \left [ 1 - \left(\frac{T_K}{T}
\right)^2 \right ]$ & $\alpha \left [ 1 -
\left(\frac{T_K}{T_v-T_K} \right)^2  \left(\frac{T_v}{T}-1
\right)^2 \right ]$
\\

& &  \\

$m_{_S}$ & $\frac{T_g^2+T_K^2}{T_g^2-T_K^2}$ &
$\frac{(T_g^2+T_K^2)-2T_gT_K (T_g/T_v)
}{(T_g^2-T_K^2)-2T_K(T_g-T_K)(T_g/T_v)}$
\\

& &  \\

$\ln{\left( \eta(T)/\eta_\infty \right )}$ & ${\frac{D
T_K}{T-T_K}\frac{T}{T+T_K} }$ & ${\frac{D
T_K}{T-T_K}\frac{T}{T+T_K-2T(T_K/T_v)} \left [
\frac{T_v-T_K}{T_v} \right]^2 } $
\\

& &  \\

\hline
\end{tabular}
\caption{Summary of the main relations relating the relevant
quantities (left column) for the gaussian configurational entropy
models: the simple gaussian model ($T_v=\infty$) and the gaussian
model with energy dependent vibrational entropy.}
\end{table*}

\section{fragility and number of states}

In the following sections we will discuss the possibility to
predict the fragility of a system from the knowledge of the
parameters characterizing the distribution of the minima of the
PEL. First, we analyze the recent works that have attempted to
relate the fragility to the "number of states". Secondly we will
see how -given a fixed configurational entropy model- one can obtain
the whole range of fragilities, thus demonstrating that, in order
to asses the fragility of a system, some additional information
is needed.

\subsection{Speedy's expression of fragility}

In 1999 Speedy \cite{Speedy99} -working in the framework of the
gaussian model and assuming the validity of the Adam-Gibbs
relation- choose to express $m_{_{S}}$ ("$f$" in his language) in
term of $\alpha$ and $\Sigma(T_g)$ ("$\Delta^l_gS(T_g)$" in
Ref.~\cite{Speedy99}). With these variables, Eq.~\ref{ms1}
becomes:
\begin{equation}
m_{_{S}} = \frac{2\alpha}{\Sigma(T_g)/Nk_B}-1. \label{ms3b}
\end{equation}
Speedy used this relation to state that "..this quantifies the
Angell observation that fragile liquids sample more basins in
configuration space than strong liquids". Actually, Eq.~\ref{ms3b}
does not help much in establishing whether or not the reported
Angell observation is correct. Indeed the proportionality between
$m_{_{S}}$ and $\alpha$ holds only if one neglects the possibility
that $\Sigma(T_g)$, a system-dependent quantity, depend on
$\alpha$. In principle its implicit dependence on $\alpha$ can
also reverse the fragility-number of states relation.

\subsection{Sastry's expression of fragility}

More recently another expression for the fragility in term of the
PEL features was derived by Sastry \cite{Sastry01}. Also in this
case the gaussian model and the Adam-Gibbs equation are at the
basis of the theory. However, Sastry does not use Eq.~\ref{relST}
to obtain the fragility. He assumed i) the validity of the VTF
law, so to relate (compare Eq.~\ref{AG} and \ref{VTF}) the
configurational entropy to the coefficient $D$, which, as
discussed before, is an index of kinetic fragility (actually,
Sastry reports his expression for the fragility $K=1/D$), and ii)
the coincidence of $T_o$ with $T_K$. The Sastry expression  takes
also into account the possible energy-depth dependence of the
basin vibrational free energy. In order to compare the expression
reported in Ref.~\cite{Sastry01} with Eq.s~\ref{ms1} and
\ref{ms3}, however, we can put the quantity $\delta S$ (in
Sastry's notation) equal to zero. The Sastry expression becomes
(with the change of notation from $\sigma$ to $\bar \epsilon$):
\begin{equation}
K = \frac{\bar \epsilon \sqrt{\alpha}}{2 \cal{E}} \left(
1+\frac{T_K}{T_g} \right ). \label{mk3}
\end{equation}
Here "$T_g$" is the MD glass transition temperatures. In
Eq.\ref{mk3} we have explicitly included the Adam-Gibbs constant
$\cal{E}$ which was implicitly assumed constant and landscape
independent in Ref.~\cite{Sastry01}(see also
\cite{nota4}).

After the conversion from $K$ to $m_{_{S}}$, using Eq.~\ref{D}, we
have:
\begin{equation}
m_{_{S}} = 17 \ln(10) \frac{\bar \epsilon \sqrt{\alpha}}{2\cal{E}}
\left( 1+\frac{T_K}{T_g} \right )+1 \label{ms4}
\end{equation}
Similarly to Eq.~\ref{ms3b}, also this equation cannot be used to
predict the $\alpha$ dependence of the fragility. Indeed, $\alpha$
appears here explicitly but also implicitly, via the
system-dependent quantities $\bar \epsilon$ and $T_K$ (see Table
I). Finally, we want to stress that the approach followed in the
derivation of the previous expression of the fragility is
intrinsically inconsistent. Indeed, as previously pointed out, the
gaussian landscape (i), the VTF law (ii) and the Adam-Gibbs
relation (iii) are not mutually consistent and, as also noticed by
Sastry \cite{Sastry01}, the hypothesis i)-iii) can only be
consistent if one uses a low temperature expansion of $\Sigma(T)$.

\subsection{Can the fragility be derived entirely \\
from the configurational entropy?}

\begin{figure}[ht]
\centering
\includegraphics[width=.43\textwidth]{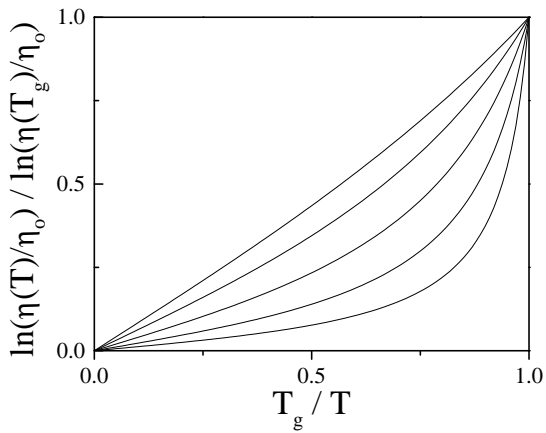}
\vspace{-5.1cm} \caption{Reconstructed Angell plot
($[\log(\eta(T)/\eta_\infty)]/[\log(\eta(T_g)/\eta_\infty)]$ vs.
$T_g/T$) for the case of the gaussian model. The different curves
correspond to different $\cal{E}$ values: from top to bottom
$\cal{E}$=80, 40, 20, 10 and 5 in units of $\sqrt{\alpha} N
\bar\epsilon/2$.} \label{f8}
\end{figure}

We aim now to prove with an example that, in general, the
configurational entropy alone is not sufficient to determine the
fragility of a system. We will use the gaussian model for the
configurational entropy and, with the help of Eq.~\ref{VTF3}, we
will set-up an "Angell plot". We could have selected any other
landscape model, reaching the same conclusion. Let's suppose to
have an hypothetical system, fully defined by a gaussian landscape
with a given value of the relevant parameters $\alpha$, $\epsilon$
and $e_o$. The temperature dependence of the viscosity in this
model is reported in Eq.~\ref{VTF3}. To set up an Angell plot, we
need to define the "glass transition temperature" $T_g$.
As done experimentally , once the $T$-dependence of the viscosity
is known, $T_g$ is defined from the condition
$\log(\eta(T_g)/\eta_\infty)=17$. Using Eq.~\ref{VTF3}, the
solution of this equation for (positive) $T_g$ is:
\begin{equation}
T_g=T_K \left \{\frac{1}{2} \frac{D}{17\ln(10)} + \sqrt{1 +
\frac{1}{4} \left( \frac{D}{17 \ln(10)} \right )^2 } \right
\}\label{xxx}
\end{equation}
with $D={\cal{E}}/(\alpha N k_B T_K)={2\cal{E}}/(\sqrt{\alpha} N
\bar\epsilon)$. For sake of compactness, let us define the
function $\gamma(x)$:
\begin{equation}
\gamma(x)=\frac{1}{2} \frac{x}{17\ln(10)} + \sqrt{1 + \frac{1}{4}
\left( \frac{x}{17 \ln(10)} \right )^2 } \label{zzz}
\end{equation}
so that
\begin{equation}
T_g=T_K\gamma(D). \label{zzzz}
\end{equation}
Obviously, the expression of $T_g$, besides the trivial
temperature scale $T_K$, depends on the parameter
$D$($={2\cal{E}}/(\sqrt{\alpha} N \bar\epsilon)$) that, in turn,
embodies the information on the "number of states" but also from
quantities distinct from the statistic of the minima
(specifically from
the parameter $\cal{E}$). We want now to plot the re-scaled
logarithmic viscosity $y(T) \equiv
[\log(\eta(T)/\eta_\infty)]/[\log(\eta(T_g)/\eta_\infty)]$ as a
function of $T_g/T$. The quantity $y(T)$, by definition of $T_g$,
turns out to be equal to $y(T)=[\log(\eta(T)/\eta_\infty)]/17$,
or, by using the expression for $\eta(T)$ reported in
Eq.~\ref{VTF3}, to:
\begin{equation}
y(T)=\frac{D}{17 \ln(10)}\frac{T_K}{T-T_K}\frac{T}{T+T_K}.
\label{y1}
\end{equation}
We can now eliminate $T_K$ from this equation in favor of $T_g$
using Eq.~\ref{zzzz} ($T_K=T_g /\gamma(D)$) to get:
\begin{equation}
y(T)=\left ( \frac{T_g}{T} \right )
\frac{\gamma(D)^2-1}{\gamma(D)^2-\left ( \frac{T_g}{T}\right )^2}.
\label{y2}
\end{equation}
In Fig.~\ref{f8} we have reported the quantity $y(T)$ of
Eq.~\ref{y2} vs. $T_g/T$, i.~e. we have made an Angell plot, for
different values of the parameter $\cal{E}$ at fixed $\alpha$ and
$\bar \epsilon$. The fragilities $m_s$ are the slopes of these
curves at the upper right corner of the plot. What is remarkable
here is that, by varying the quantity $\cal{E}$ entering in the
numerator of the exponent in the Adam-Gibbs relation (Eq.[9]) at
fixed configurational entropy, we can span the whole range of
fragilities. In other words, for a given (gaussian in the present
example) landscape, with well defined statistical properties
(fixed $\alpha$ and $\bar\epsilon$), we can have a strong system
(large $\cal{E}$) as well as a fragile one (small $\cal{E}$).
Therefore, we conclude this section with the statement that in
principle -whenever the Adam Gibbs relation represents a good
approximation of the relation between transport properties and
configurational entropy-  {\it the knowledge of the
configurational entropy alone would be not sufficient to define
the fragility of a system}\cite{note}. This statement, and the
role of the effective barrier height in determining the fragility
of a glass has been already discussed in literature (see e.~g.
Ref.~\cite{Perera}). The previous conclusion does not imply that
the fragility cannot be derived from the landscape properties:
indeed, it is possible, and actually most likely, that the
quantity $\cal{E}$ could be derived from other features of the PEL
than the minima distribution, as, for example the
minimum-to-minimum barrier heights. Future studies must focus on
the relation between $\cal{E}$ and the PEL properties and on the
physical range of values of ${\cal{E}}$.

\subsection{Strong-to-fragile transition}

\begin{table*}[t]
\begin{tabular}{@{\hspace{15pt}} l @{\hspace{15pt}} |*{4}{@{\hspace{20pt}} c}}
\hline
& & & & \\
& $T_g$ & $T_K$  & $m_{_A}$ & Ref. \\
& & & & \\
\hline
& & & & \\
2-metylpentane        & 80.5 &   58 & 58   & \cite{Angell97},\cite{Angell97} \\
butyronitrile         & 100  & 81.2 & 47   & \cite{Angell97},\cite{Angell97} \\
ethanol               & 92.5 &   71 & 55   & \cite{Angell97},\cite{McKenna} \\
n-propanol            &102.5 &   73 & 36.5 & \cite{Angell97},\cite{Angell97} \\
toluene               & 126  &   96 & 59   & \cite{Angell97},\cite{McKenna} \\
1-2 propan diol       & 172  &  127 & 52   & \cite{Angell97},\cite{Angell97} \\
glycerol              & 190  &  135 & 53   & \cite{Angell97},\cite{Angell97} \\
triphenil phospate    & 205  &  166 & 160  & \cite{Angell97},\cite{Angell97} \\
orthoterphenyl        & 244  &  200 & 81   & \cite{Angell97},\cite{Angell97} \\
m-toluidine           & 187  &  154 & 79   & \cite{Angell97},\cite{Angell97} \\
propylene carbonate   & 156  &  127 & 104  & \cite{Angell97},\cite{Angell97} \\
sorbitol              & 266  &  226 & 93   & \cite{Angell97},\cite{Angell97} \\
selenium              & 307  &  240 & 87   & \cite{Angell97},\cite{Angell97} \\
ZnCl$_2$              & 380  &  250 & 30   & \cite{Angell97},\cite{McKenna} \\
As$_2$S$_3$           & 455  &  265 & 36   & \cite{Angell97},\cite{McKenna} \\
CaAl$_2$Si$_2$O$_8$   &1118  &  815 & 53   & \cite{Angell97},\cite{McKenna} \\
Propilen Glycol       & 167  & 127  & 52   & \cite{Richert98},\cite{Richert98}
\\
3-Methyl pentane      &  77  & 58.4 & 36   & \cite{Richert98},\cite{Richert98}
\\
3-Bromopentane        & 108  & 82.5 & 53   & \cite{Richert98},\cite{Richert98}
\\
2-methyltetrahydrofuran &  91  & 69.3 & 65   & \cite{Richert98},\cite{Richert98}
\\
& & & & \\
\hline
\end{tabular}
\caption{Summary of the quantity relevant to test Eq.~\ref{ms1}
($T_g$, $T_K$ and $m_{_S}$) for those systems where the three
quantity are all known. The last column reports the references
where the data has been found. In those case where more than one
value of the parameters are known, we have reported here the
average value. The first of the two references refers to the
couple ($T_g$, $T_K$) and the second to $m_{_S}$.}
\end{table*}

\begin{figure}[ht]
\centering
\includegraphics[width=.43\textwidth]{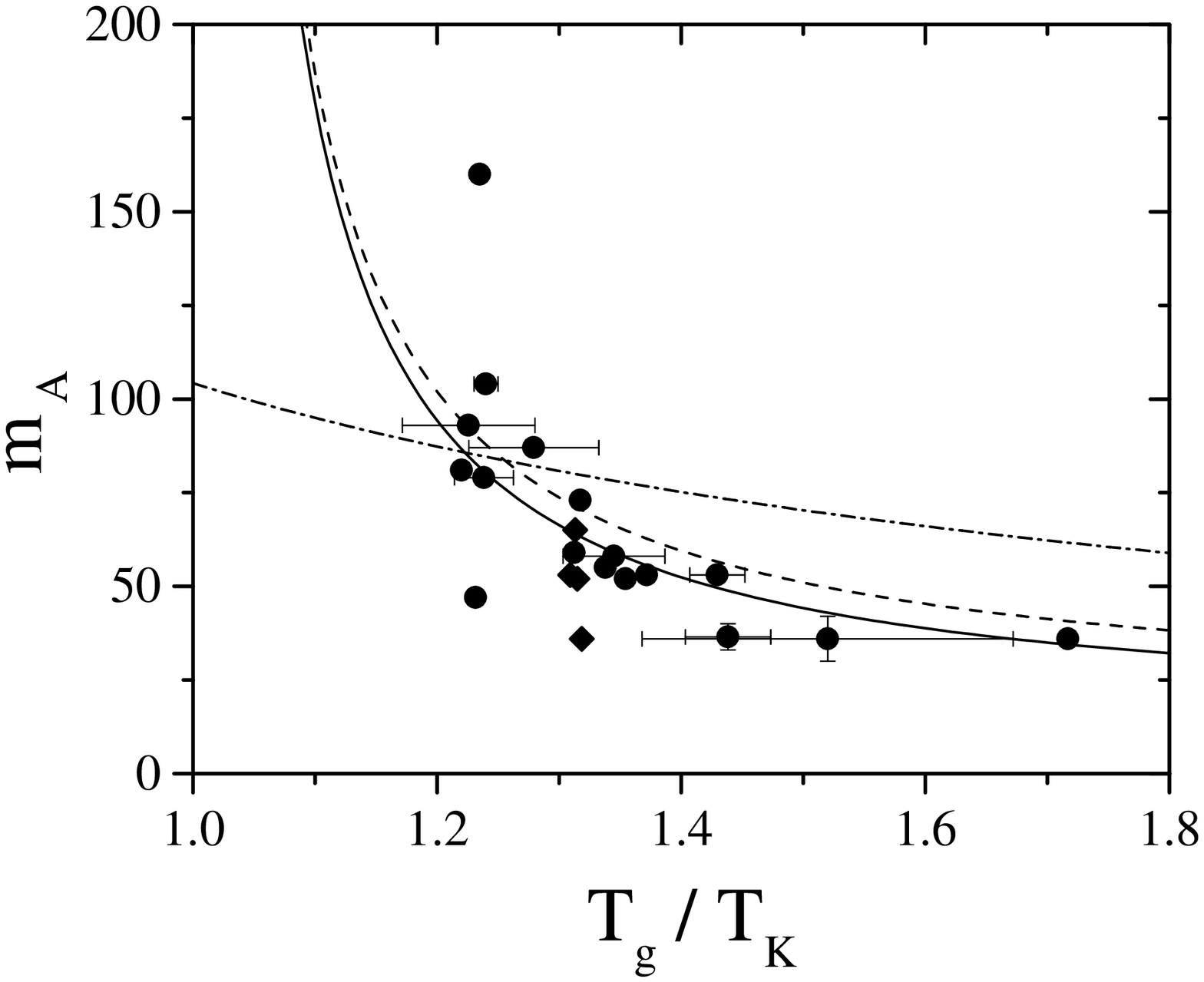}
\vspace{-5.1cm} \caption{Experimental values of the kinetic
fragility $m_{_A}$ plotted as a function of the ratio $T_g/T_K$
for those systems where the three quantities ($m_{_A}$, $T_g$,
and $T_K$) are available. The input data are reported in Table
III. For those systems where more than one determination of the
parameters is known, we have reported in the plot the average
value together with an "error" bar that indicates the whole
dispersion.} \label{f9}
\end{figure}

In the previous section we have shown that, on a general ground, a
simple gaussian landscape with fixed statistical properties could
be shared by the whole class of known systems; they would simply
differ in the value of $\cal{E}$ that, in turn, induces a
different value of $T_g/T_K$, thus a different fragility. In this
scheme a fragile system -having $T_g$ close to $T_K$ (as deduced
from Eq.~\ref{ms1})- visits that part of the landscape where
$\Sigma(e)$ is strongly $e$ dependent, thus (see Eq.~\ref{relST})
pushing $m_{_S}$ up. On the contrary, a strong system has $T_g$
far away from $T_K$, and the system is confined to visit the
region where $\Sigma(e)$ is almost flat. In other words, if all
the system shared the same landscape, due to the difference in the
parameter $\cal{E}$, a strong system (large $\cal{E}$) would visit
the "top-of- the-landscape", while a fragile system (small
$\cal{E}$) would be allowed to go down in energy. If this scenario
were correct, we would expect that real systems verify
Eq.~\ref{ms1} (or \ref{ms2}, or \ref{ms3}). In Fig.~\ref{f9} we
report, for those systems where all the three quantities $T_g$,
$T_K$ and $m_{_A}$ are known (see Table III), the fragility
$m_{_A}$ as a function of $T_g/T_K$ (symbols). Also shown in the
same figure are the predictions of Eq.s~\ref{ms1}, \ref{ms2},
\ref{ms3} (lines). Few points must be underlined: i) there is a
rather good general agreement, but the single systems does not
strictly verifies none of the three predictions. This can be due
to the existence of landscapes different from the three simple
cases discussed at the beginning of this chapter, or, most likely,
to the presence of a finite value for $T_v$. Indeed, recent
molecular simulations of model liquids \cite{simTV} clearly show
such a phenomenology, indicating a non negligible energy
dependence of the vibrational entropy. ii) The differences among
Eq.~\ref{ms1} and \ref{ms2} are so small, that the experimental
data do not allow to discriminate among these two different
landscape models, while the (pure) logarithmic model seems to be
definitively unacceptable. Most likely, a gaussian model with a
small logarithmic correction would be still acceptable. iii) Among
the systems represented in Fig.~\ref{f9} the lowest fragility is
$\approx$35, i.~e. the strong systems are absent (for these
systems a reliable estimation of $T_K$ does not exist), and this
does not allow to firmly establish the general validity of one of
the three model, and, more generally, of the idea presented before
that strong systems and fragile systems are characterized by a
common configurational entropy and a different elevation in the
PEL.

Of course, we are not stating that the depicted behaviour is the
actual one. Different systems have different value of $\alpha$ and
may even not be described by a common landscape model. A typical
example, that it is worth to discuss here, is the case of silica.
As shown by Horbach and Kob \cite{Horbach99}, vitreous silica (as
described by the BKS potential model \cite{BKS}) show strong
$T$-dependence of the fragility. More specifically, v-$SiO_2$,
which is a well known strong system close to the glass transition
temperature, turns toward a more fragile behaviour on increasing
$T$. This phenomenon, called "strong-to-fragile" transition, first
proposed for the case of water\cite{watStF1}, has been observed in
simulations of  water \cite{watStF} and  (simulated) berillium
fluoride \cite{BeF2StF}. It is obvious that the fragile-to-strong
transition cannot be framed within the possibility described in
the first paragraph of this section.

In a recent simulations work, Saika-Voivod, Sciortino and Poole
\cite{SSP} have shown that the configurational entropy for liquid
silica -as derived from a MD simulation based on the BKS
interaction potential model- is far from being "gaussian". More
specifically, they found that $\Sigma(T)$ -at low $T$- shows a
tendency towards a positive curvature and does not seem to
extrapolate to zero entropy at a finite temperature. This behavior
is shared by the logarithmic model (or by a combination of the
logarithmic and gaussian models, as proposed in
Ref.~\cite{DebStiShell}) for $\Sigma(e)$. This model predicts an
infinite slope of $\Sigma(e)$ at $e_K$, and this could be in
agreement with the simulation results of Ref.~\cite{SSP} as the
low statistics in the tail of $\Omega_N(E)$ -as measured by MD-
does not allow to safely determine $d\Sigma(e)/de$ evaluated at
$e_K$.

\begin{figure}[ht]
\centering
\includegraphics[width=.43\textwidth]{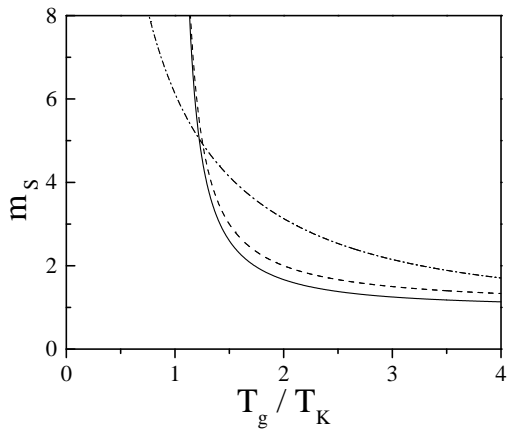}
\vspace{-5.1cm} \caption{Temperature dependence of the fragility
for the three examined models: gaussian (full line), hyperbolic
(dashed line) and logarithmic(dot-dashed line). The quantity
$m_{_S}$ is reported as a function of $T_g/T_K$.} \label{f10}
\end{figure}

The logarithmic model, however, similarly to the other models
presented before, is not capable to catch the physics of the
strong-to-fragile transition. Indeed, the fragility expressions
for all the examined models (Eq.s~\ref{ms1}, \ref{ms2}, \ref{ms3})
show a {\it monotonic} $T$ dependence, whit a tendency toward a
{\it decrease} of the fragility on increasing temperature (see
Fig.~\ref{f10}). A behavior opposite to what observed in simulated
vitreous silica. It is therefore clear that an infinite slope of
$\Sigma(e)$ at $e_K$ alone is not sufficient to guarantee the
existence of a strong-to-fragile transition. What is actually
sufficient (necessary ?) for a strong to fragile transition -i.~e.
to have a maximum in the $m_{_S}$ vs. $T_g/T_K$ function- is that
the configurational entropy -as a function of $T$- had a non-zero
limit for $T \rightarrow 0$. This can be understood looking at
Eq.~\ref{relST}. It is clear that a fragile system is
characterized by a large value of $\Sigma'(T)$ (fragile systems
explore the "steep" part of the PEL), while a strong system will
have a small value of $\Sigma'(T)$, but also a non zero
$\Sigma(T)$. This certainly happens at the "top of the landscape",
but could also happen at low $T$ if $\Sigma(0) \neq 0$ (in the
logarithmic model, at low $T$, $\Sigma'(T) \rightarrow 0$, but the
same does $\Sigma(T)$ and the resulting fragility increases
continuously). Thus,  a strong-to-fragile transition could take
place only if the landscape of the systems allows for a finite
number of states at zero temperature, i.~e. for a (exponentially
large with $N$) degenerate fundamental state. The existence of
such a degeneracy for system with short range interaction (non
mean field systems) poses several problems (see the discussion in
Ref.~\cite{DebStiShell}), and is certainly calling for further
investigation.

\section{Discussion and Conclusion}

In conclusion, in this paper we have first summarized the main
definition of fragility, then we have recalled and studied
different models for the configurational entropies present in the
literature. Using the Adam-Gibbs relation to link the dynamics of
a glass forming system to its configurational entropy, we have
reported the explicit expressions for different quantities, among
which the fragility. From the reported relation, it is clear that
in general {\it the fragility cannot be derived by the knowledge
of the configurational entropy}. More specifically, given a fixed
"landscape", different system fragility can be mimicked by
varying the parameter $\cal E$ entering in the numerator of the
exponent of the Adam-Gibbs equation. On a general ground, the
fragility of a system depends on the ratio ${\cal E}/\alpha N k_B
T_K$.

The fact that the whole range of fragility can be derived from a
given PEL model (e.~g. the gaussian model) with the same
statistical properties seems an interesting possibility. If this
was the case, the strong glass forming materials would be
characterized by a large value of $\cal E$ and would explore the
"top-of-the-landscape", while the most fragile ones would have
small $\cal E$ and would visit the states around the inflection
point of $\Sigma(T)$. Obviously other possibilities exist, as for
example that all the systems were characterized by the same $\cal
E$, and in this case strong glass would have a small number of
states (small $\alpha$), at variance to the fragile systems with
more states (large $\alpha$). A further scenario can be
hypothesized, that would also explain the existence of a
strong-to-fragile transition: in this case the strong systems
would explore the bottom of a landscape characterized by a
non-vanishing zero point entropy. This is an interesting
possibility that deserve deeper investigation.

Overall, the present discussion, which heavily build on the
validity of the Adam and Gibbs relation, indicates that in
principle at least two possible classes of strong glass forming
materials can actually exist. On one side we have those systems
that -close to $T_g$- visit state at the top of the landscape and
have a "regular" (gaussian-like) configurational entropy (let's
call these systems as class A strong glass forming materials). On
the other side we find the -let's say- class B strong liquids,
that visit minima deep in the PEL, but with an exponentially large
degeneracy of the fundamental state. The answer to the question
whether class A and/or B strong systems actually exist requires
further investigations.

As a final comment, we would like to recall that fragility is
often measured at \textit{constant pressure}, while all the
configurational entropy based models -as those presented here- are
built on the assumption of a well defined $\Sigma(e)$
function,i.~e. they assume \textit{constant density}. The
relationship between constant density and constant pressure
fragilities is one of the topic under discussion at the present
time. As an example, in the case of soft sphere systems it has
been shown \cite{S5} that $\Sigma_c(T)$ along isochoric and
isobaric paths are very close one to each other. Similarly, in a
very recent work \cite{Tarjus03}, Tarjus and coworkers show that
-in alcohols- the change in density only slightly affects the
fragility, thus indicating that under the experimentally
accessible density changes the landscape suffers only  minor
modifications. For other systems, on the contrary, it has been
observed a large deviation ($\approx 40 \%$) between constant
density and constant pressure fragilities \cite{Sichina}. This
ongoing discussions, however, does not affect the conclusions of
the present work since all the formalism could have been based on
the {\it enthalpy} landscape, instead of {\it energy} landscape,
without any changes in the results.

\section{Acknowledgment}
We acknowledge very useful discussions with G.~Parisi and
S.~Sastry. We also thanks L.~Angelani, J.~Dyre, S.~Mossa, K.~Ngai,
S.~Sastry and R.~Speedy for a critical reading of the manuscript
and for helpful suggestions. A special thanks to Austen Angell for
his invaluable suggestions that helped us in the preparation of
this manuscript. We acknowledge support from Miur FIRB and Cofin
and INFM.


\begin{thebibliography}{99}
\bibitem{nnn}
The total number of inherent structure of the PEL is the product
of $\Omega_N$ times the trivial factor $N!$ due to particles
permutation.
\bibitem{nota2}
A remarkable exception is represented by vitreous silica, which is
one of the strongest systems according to the kinetic fragility
while turns out very close to glycerol (intermediate value of
$m_{_{A}}$) as far as the $T$-dependence of the excess entropy is
concerned.
\bibitem{nota4}
Note that Eq.~8 in Ref.\cite{Sastry01} defines a fragility index
with the dimension of  energy. In the caption of Fig.~1, Sastry
clarifies that --- in order to have a dimensionless fragility
index --- a dimensional quantity is needed in going from from
$TS_c$ which is a dimensional quantity and fragility which is
dimensionless.  In the absence of a good way to calculate ${\cal
E}$, Sastry choose the Lennard Jones energy scale $\epsilon_{AA}$,
finding support for such a choice in the numerical data. This
choice leads to the approximation $K = \frac{\bar \epsilon
\sqrt{\alpha}}{2 \varepsilon_{AA}} \left( 1+\frac{T_K}{T} \right
)$, to be compared with the exact expression $K = \frac{\bar
\epsilon \sqrt{\alpha}}{2 \cal{E}} \left( 1+\frac{T_K}{T} \right
)$, reported in Eq.~\ref{mk3}.
\bibitem{note}
The existing correlation between thermodynamic ($m_T$) and kinetic ($m_s$)
fragilities -as experimentally observed in Ref. 32-
 is NOT in contradiction with
the statement that $\Sigma(T)$ alone is not sufficient to determine fragility.
Indeed, the experimentally observed correlation between $m_T$ and $m_s$
(discussed in Eq. 11 of the manuscript) can be derived ---beyond assuming the
proportionality between  ÒconfigurationalÓ and ÒexcessÓ entropy ---   from \\i)
the definition of $m_T$ (Eq. 7),\\  ii) the definition of $m_s$ (Eq. 2), \\iii)
the AG  relation\\We stress that Eq. 11 is independent from whether or not the
fragility could be derived from $\Sigma(T)$ only. Note that, as shown in Eq. 7,
the evaluation of $m_T$ requires the knowledge of two thermodynamic quantities
(the excess entropy and the T-derivative of it) but also the knowledge of one
dynamical  quantity ($T_g$). The presence of $T_g$ in the definition of $m_T$
highlight the need of dynamical information in the evaluation of the fragility.
\end{thebibliography}
\end{document}